\documentclass{article}
\usepackage[T1]{fontenc}
\usepackage[latin9]{inputenc}
\usepackage{amsmath}
\usepackage{amsthm}
\usepackage{amssymb}

\makeatletter
\numberwithin{equation}{section}
\numberwithin{figure}{section}

\@ifundefined{date}{}{\date{}}
\makeatother

\usepackage[english]{babel}
\begin{document}
\begin{flushleft} {\bf OUJ-FTC-10} \end{flushleft}
\begin{center} 

{\LARGE{}Unlocking Novel Quantum States: Virasoro-Bogoliubov Transformations
in Two Modes}{\LARGE\par}
\vspace{32pt} 

So Katagirii\textsuperscript{{*}}\footnote{So.Katagiri@gmail.com}
\vspace{10pt}

\begin{center}
\textit{Division of Arts and Sciences, The School of Graduate Studies,
The Open University of Japan, Chiba 261-8586, Japan}
\par\end{center}

\vspace{10pt}

\end{center}

\begin{abstract}
This paper explores the Bogoliubov transformation's extension to two-mode
squeezed states, building on our previous work with Virasoro-squeezing.
We establish the Virasoro-Bogoliubov transformation as a non-linear
extension of the traditional Bogoliubov transformation, creating non-linear
two-mode squeezed states. This research unveils novel quantum states
with the potential for innovative insights in various fields of quantum
physics.
\end{abstract}

\section{Introduction}

The Bogoliubov transformation is a fundamentally important physical
tool in quantum field theory. This transformation essentially exhibits
non-perturbative effects, providing insights into physics that are
unreachable through perturbation theory. It stands as a core idea
in the BCS theory of superconductivity\cite{key-19} and was led into
discussions of spontaneous symmetry breaking in particle physics by
the Nambu theory\cite{key-20}. Furthermore, it proves crucial in
discussions of Hawking radiation\cite{key-21} and Rindler spacetime
in black hole physics\cite{key-22,key-23,key-24}. Discourses on the
further generalization of such states have been advanced in the field
of quantum optics\cite{key-25}.

Generalizing the Bogoliubov transformations will expand the range
of physical phenomena described by these various Bogoliubov transformations
to more complex cases, and also has the potential to provide new insights
by viewing these phenomena from a more general perspective. In addition,
nonlinear generalizations of Bogoliubov transformations open up the
possibility of new physical properties and applications to quantum
computing as new higher-order entanglement states. 

In our previous research, we proposed a methodology called Virasoro-squeezing,
a generalization of squeezing from the perspective of the Virasoro
algebra, and constructed non-linear single-mode squeezed states\cite{key-26}.

In this study, we extend this to two modes, establishing the Virasoro-Bogoliubov
transformation as a non-linear extension of the Bogoliubov transformation
using the Virasoro algebra. As a result, the Bogoliubov transformation
generalizes nonlinearly as a squeezing of the center of mass coordinates
and relative coordinates.

The structure of this paper is as follows: 

In Section 2, we will provide an explanation about the Virasoro algebra.
In Section 3, we will review our previous paper in which we constructed
Virasoro-squeezing. In Section 4, we reconsider from the perspective
of the relationship between Bogoliubov transformations and squeezing.
In Section 5, we generalize the Bogoliubov transformation from the
perspective of Virasoro squeezing. In the final section, we summarize
our conclusions and carry out discussions.

\section{Virasoro algebra}

A two-dimensional Virasoro algebra is an infinite-dimensional algebra
with scale transformations, whose generators satisfy the following
algebraic relations:

\begin{equation}
[L_{n},\ L_{m}]=(n-m)L_{n+m}+\frac{c}{12}(n^{3}-n)\delta_{m+n,0},
\end{equation}
where, $c$ is called the central charge and commutes with any $L_{n},\ [c,\ L_{n}]=0$.

$L_{-1},L_{0},\ L_{1}$ satisfies

\begin{equation}
[L_{\pm1},\ L_{0}]=\pm L_{\pm1},\ [L_{1},L_{-1}]=2L_{0}
\end{equation}
which is $\mathrm{SL}(2,\mathbb{C})$ subalgebra of Virasoro algebra.

When c = 0, this algebra reduces to

\begin{equation}
[L_{n},\ L_{m}]=(n-m)L_{n+m},
\end{equation}
which is called Witt algebra (centerless Virasoro algebra) \cite{key-5}.

The generator of this algebra Ln can be constructed by z and its differential
operator $\partial$ as follows:

\begin{equation}
L_{n}=z^{n+1}\partial
\end{equation}

Then, we can understand the specific geometrical meaning of these
algebras.

\begin{equation}
L_{0}=z\partial
\end{equation}
generates a scale transformation,
\begin{equation}
e^{\theta L_{0}}f(z)=f(e^{\theta}z).
\end{equation}

Next,
\begin{equation}
L_{-1}=\partial
\end{equation}
generates a parallel transformation,
\begin{equation}
e^{\theta L_{-1}}f(z)=f(z+\theta).
\end{equation}

Finally,
\begin{equation}
L_{1}=z^{2}\partial
\end{equation}
generates a special conformal transformation,

\begin{equation}
e^{\theta L_{1}}f(z)=f\left(\frac{z}{1-\theta z}\right).
\end{equation}

As mentioned above, Virasoro algebra (or Witt algebra) is a more general
algebra that includes scale transformations and translations.

\section{Virasoro Squeezed State}

Squeezing transformations are defined by the following generators:

\begin{equation}
\hat{G}=\frac{1}{2}(\hat{a}^{2}-\hat{a}^{\dagger2}).
\end{equation}
using the creation and annihilation operators $\hat{a}^{\dagger},\hat{a}$.

The creation and annihilation operators transform by this transformation
as follows: 

\begin{equation}
\hat{a}(\theta)=\hat{U}(\theta)\hat{a}\hat{U}^{\dagger}(\theta)=\hat{a}\cosh\theta+\hat{a}^{\dagger}\sinh\theta,
\end{equation}

\begin{equation}
\hat{a}^{\dagger}(\theta)=\hat{U}(\theta)\hat{a}^{\dagger}\hat{U}^{\dagger}(\theta)=\hat{a}^{\dagger}\cosh\theta+\hat{a}\sinh\theta,
\end{equation}

where $\hat{U}(\theta)=e^{\theta\hat{G}}$.

The characteristics of squeezing are revealed in the transformation
of position and momentum operators.

The position and momentum operator are constructed as 
\begin{equation}
\hat{x}=\sqrt{\frac{1}{2\omega_{0}}}\left(\hat{a}+\hat{a}^{\dagger}\right),
\end{equation}

\begin{equation}
\hat{p}=i\sqrt{\frac{\omega_{0}}{2}}\left(\hat{a}-\hat{a}^{\dagger}\right).
\end{equation}

from the creation and annihilation operators. From this, the generator
becomes 
\begin{equation}
G=i(\hat{x}\hat{p}+\hat{p}\hat{x}).
\end{equation}

Position and momentum are transformed by the Squeezing transformation
as 

\begin{equation}
\hat{x}(\theta)=\hat{U}(\theta)\hat{x}\hat{U}^{\dagger}(\theta)=e^{\theta}\hat{x},
\end{equation}

\begin{equation}
\hat{p}(\theta)=\hat{U}(\theta)\hat{p}\hat{U}^{\dagger}(\theta)=e^{-\theta}\hat{p}.
\end{equation}

Thus, squeezing is a transformation that stretches one side of the
phase space (position in the current case) and contracts the other
side (momentum in the current case).

Squeezed state is given by

\begin{equation}
|\theta\rangle=\hat{U}(\theta)|0\rangle=e^{\theta\hat{G}}|0\rangle.
\end{equation}

Virasoro-Squeezing is a natural extension to Virasoro algebra of the
fact that squeezing can be regarded as a combination of scale transformations.

Its generator is introduced as 

\begin{equation}
\hat{L}_{n}\equiv-\frac{i}{2}\left(\hat{x}^{n+1}\hat{p}+\hat{p}\hat{x}^{n+1}\right).
\end{equation}

$\hat{L}_{n}$ satisfies

\begin{equation}
[\hat{L}_{n},\hat{L}_{m}]=(n-m)\hat{L}_{n+m},
\end{equation}
This is the centerless Virasoro algebra (Witt algebra). We note $\hat{L}_{0}$
is a generator of usual second-order squeezing transformation.

Position and momentum are transformed by the Virasoro-Squeezing transformation
is given 

\begin{equation}
\hat{x}(\theta)=\hat{S}_{n}(\theta)\hat{x}\hat{S}_{n}^{\dagger}(\theta)=(1+n\theta\hat{x}^{n})^{-\frac{1}{n}}\hat{x},
\end{equation}

\begin{equation}
\hat{p}(\theta)=\hat{S}_{n}(\theta)\hat{p}\hat{S}_{n}^{\dagger}(\theta)=\frac{1}{2}\{(1+n\theta\hat{x}^{n})^{+\frac{1}{n}+1},\hat{p}\}.
\end{equation}

Then, Virasoro squeezed state is given by

\begin{equation}
|\theta\rangle_{n}\equiv\hat{S}_{n}|0\rangle=e^{\theta\hat{L}_{n}}|0\rangle.
\end{equation}

The expected number of particles in the Virasolo Squeezed state is 

\begin{equation}
_{n}\langle\theta|\hat{N}|\theta\rangle_{n}=\langle0|\left(\sinh\log(1+n\theta\hat{x}^{n})^{-\frac{1}{n}-\frac{1}{2}}\right)^{2}\hat{a}\hat{a}^{\dagger}|0\rangle,
\end{equation}
which coincides with the expected number of particles in the normal
Virasolo state, 

\begin{equation}
\langle\theta|\hat{N}|\theta\rangle=\left(\sinh\theta\right)^{2},
\end{equation}
when $n=0$.

\section{Bogoluibov transforamtion and Squeezing}

The Bogoluibov transformation is defined by a generator like 

\begin{equation}
\hat{G}=\hat{a}_{1}\hat{a}_{2}-\hat{a}_{1}^{\dagger}\hat{a}_{2}^{\dagger},
\end{equation}
using the creation and annihilation operators $\hat{a}_{1}$ and $\hat{a}_{2}$
corresponding to the two modes.

The creation and annihilation operators transform by this transformation
as follows: 

\begin{equation}
\hat{a}_{1}(\theta)=\hat{U}(\theta)\hat{a}_{1}\hat{U}^{\dagger}(\theta)=\hat{a}_{1}\cosh\theta+\hat{a}_{2}^{\dagger}\sinh\theta,
\end{equation}

\begin{equation}
\hat{a}_{2}^{\dagger}(\theta)=\hat{U}(\theta)\hat{a}_{2}\hat{U}^{\dagger}(\theta)=\hat{a}_{1}\sinh\theta+\hat{a}_{2}^{\dagger}\cosh\theta,
\end{equation}
where $\hat{U}(\theta)=e^{\theta\hat{G}}$.

The position and momentum operators are constructed as 

\begin{equation}
\hat{x}_{i}=\sqrt{\frac{1}{2\omega_{0}}}\left(\hat{a}_{i}+\hat{a}_{i}^{\dagger}\right),
\end{equation}

\begin{equation}
\hat{p}_{i}=i\sqrt{\frac{\omega_{0}}{2}}\left(\hat{a}_{i}-\hat{a}_{i}^{\dagger}\right),
\end{equation}

from the creation and annihilation operators. From this, the generator
becomes 

\begin{equation}
\hat{G}=4i(\hat{x}_{1}\hat{p}_{2}+\hat{x}_{2}\hat{p}_{1}).
\end{equation}

Position and momentum are transformed by the Bogoliubov transformation
as 

\begin{equation}
\hat{x}_{1}(\theta)=\hat{x}_{1}\cosh\theta+\hat{x}_{2}\sinh\theta,
\end{equation}

\begin{equation}
\hat{p}_{1}(\theta)=\hat{p}_{1}\cosh\theta-\hat{p}_{2}\sinh\theta.
\end{equation}

This transformation is not a squeezed transformation of position and
momentum. Therefore, it is not possible to make generalizations such
as those for Virasoro-Squeezing for each mode.

To rewrite the Bogoliubov transformation in terms of squeezing, We
 introduce the following center of mass coordinates and relative coordinates: 

\begin{equation}
\hat{X}\equiv\frac{\hat{x}_{1}+\hat{x}_{2}}{2},\ \Delta\hat{x}=\hat{x}_{1}-\hat{x}_{2}.
\end{equation}

\begin{equation}
\hat{P}\equiv\frac{\hat{p}_{1}+\hat{p}_{2}}{2},\ \Delta\hat{p}=\hat{p}_{1}-\hat{p}_{2}.
\end{equation}

These are shown to be squeezing transformations in the Bogoliubov
transformation, respectively:

\begin{equation}
\hat{X}(\theta)=e^{\theta}\hat{X},
\end{equation}

\begin{equation}
\Delta\hat{x}(\theta)=e^{-\theta}\Delta\hat{x},
\end{equation}

\begin{equation}
\hat{P}(\theta)=e^{-\theta}\hat{P},
\end{equation}

\begin{equation}
\Delta\hat{p}(\theta)=e^{+\theta}\Delta\hat{p}.
\end{equation}

The generator can be written 
\begin{equation}
\hat{G}=\hat{X}\hat{P}+\hat{P}\hat{X}-\frac{1}{4}\left(\Delta\hat{x}\Delta\hat{p}+\Delta\hat{p}\Delta\hat{x}\right)
\end{equation}

in these variables.

\section{Virasoro Bogoluibov transformation}

In the previous section, the Bogoliubov transformation could be viewed
as a squeezing of the center of mass and relative coordinates, respectively.
From this, the Vilasoro-Bogoliubov transformation 

\begin{equation}
\hat{L}_{n}\equiv\hat{X}^{n+1}\hat{P}+\hat{P}\hat{X}^{n+1}-\frac{1}{4}\left(\Delta\hat{x}^{n+1}\Delta\hat{p}+\Delta\hat{p}\Delta\hat{x}^{n+1}\right)
\end{equation}
can be constructed by generalizing the squeezing to the Vilasoro-Squeezing.

This transformation transforms the center of mass coordinates and
relative coordinates as 

\begin{equation}
\hat{X}(\theta)=\hat{S}_{n}\hat{X}\hat{S}_{n}^{\dagger}=(1+n(\theta\hat{X}^{n}))^{-\frac{1}{n}}\hat{X},
\end{equation}

\begin{equation}
\Delta\hat{x}(\theta)=\hat{S}_{n}\Delta\hat{x}\hat{S}_{n}^{\dagger}=(1+n(\theta\Delta\hat{x}^{n}))^{-\frac{1}{n}}\Delta\hat{x},
\end{equation}

\begin{equation}
\hat{P}(\theta)=\hat{S}_{n}\hat{P}\hat{S}_{n}^{\dagger}=\frac{1}{2}(1+n(\theta\hat{X}^{n}))^{\frac{1}{n}+1}\hat{P}+\frac{1}{2}\hat{P}(1+n(\theta\hat{X}^{n}))^{\frac{1}{n}+1},
\end{equation}

\begin{equation}
\Delta\hat{p}(\theta)=\hat{S}_{n}\Delta\hat{p}\hat{S}_{n}^{\dagger}=\frac{1}{2}(1+n(\theta\Delta\hat{p}^{n}))^{\frac{1}{n}+1}\Delta\hat{p}+\frac{1}{2}\Delta\hat{p}(1+n(\theta\Delta\hat{p}^{n}))^{\frac{1}{n}+1},
\end{equation}

where $\hat{S}_{n}(\theta)=e^{\theta\hat{L}_{n}}$.

From these transformations, the creation and annihilation operators
are obtained as nonlinear generalizations of the following Bogoliubov
transformations:

\begin{equation}
\begin{aligned}\hat{a}_{1}(\theta)= & \frac{1}{2}\hat{K}(n,\hat{X})\left(\cosh\Omega(n,\hat{X})(\hat{a}_{1}+\hat{a}_{2})+\sinh\Omega(n,\hat{X})(\hat{a}_{1}^{\dagger}+\hat{a}_{2}^{\dagger})\right)\\
 & +\frac{1}{2}\left((\hat{a}_{1}+\hat{a}_{2})\cosh\Omega(n,\hat{X})+(\hat{a}_{1}^{\dagger}+\hat{a}_{2}^{\dagger})\sinh\Omega(n,\hat{X})\right)\hat{K}(n,\hat{X})\\
 & +\frac{1}{4}\hat{K}(n,\Delta\hat{x})\left(\cosh\Omega(n,\Delta\hat{x})(\hat{a}_{1}-\hat{a}_{2})+\sinh\Omega(n,\Delta\hat{x})(\hat{a}_{1}^{\dagger}-\hat{a}_{2}^{\dagger})\right)\\
 & +\frac{1}{4}\left((\hat{a}_{1}-\hat{a}_{2})\cosh\Omega(n,\Delta\hat{x})+(\hat{a}_{1}^{\dagger}-\hat{a}_{2}^{\dagger})\sinh\Omega(n,\Delta\hat{x})\right)\hat{K}(n,\Delta\hat{x}),
\end{aligned}
\end{equation}

\begin{equation}
\begin{aligned}\hat{a}_{2}(\theta)= & \frac{1}{2}\hat{K}(n,\hat{X})\left(\cosh\Omega(n,\hat{X})(\hat{a}_{1}+\hat{a}_{2})+\sinh\Omega(n,\hat{X})(\hat{a}_{1}^{\dagger}+\hat{a}_{2}^{\dagger})\right)\\
 & +\frac{1}{2}\left((\hat{a}_{1}+\hat{a}_{2})\cosh\Omega(n,\hat{X})+(\hat{a}_{1}^{\dagger}+\hat{a}_{2}^{\dagger})\sinh\Omega(n,\hat{X})\right)\hat{K}(n,\hat{X})\\
 & -\frac{1}{4}\hat{K}(n,\Delta\hat{x})\left(\cosh\Omega(n,\Delta\hat{x})(\hat{a}_{1}-\hat{a}_{2})+\sinh\Omega(n,\Delta\hat{x})(\hat{a}_{1}^{\dagger}-\hat{a}_{2}^{\dagger})\right)\\
 & -\frac{1}{4}\left((\hat{a}_{1}-\hat{a}_{2})\cosh\Omega(n,\Delta\hat{x})+(\hat{a}_{1}^{\dagger}-\hat{a}_{2}^{\dagger})\sinh\Omega(n,\Delta\hat{x})\right)\hat{K}(n,\Delta\hat{x}),
\end{aligned}
\end{equation}
where 
\begin{equation}
e^{\hat{\Omega}(n,\hat{x})}\equiv\hat{K}(n,\hat{x})^{-\frac{2}{n}-1},
\end{equation}

\begin{equation}
\hat{K}(n,\hat{x})\equiv(1+n\theta\hat{x}^{n})^{\frac{1}{2}}.
\end{equation}

This is consistent with the usual Bogoliubov transformation at $n=0$.

Thus, Virasoro Bogoluibov state is given by

\begin{equation}
|\theta\rangle_{n}\equiv\hat{S}_{n}|0\rangle=e^{\theta\hat{L}_{n}}|0\rangle.
\end{equation}

Using these equations, if we take mean field, 
\begin{equation}
K(n,\hat{x})\approx K(\theta,\langle x^{n}\rangle)\equiv(1+n\theta\langle\hat{x}^{n}\rangle)^{\frac{1}{2}}
\end{equation}

\begin{equation}
e^{\hat{\Omega}(n,\hat{x})}\approx e^{\Omega(\theta,\langle\hat{x}^{n}\rangle)}\equiv K(\theta,\langle\hat{x}^{n}\rangle)^{-\frac{2}{n}-1}
\end{equation}

We obtain,
\begin{equation}
\begin{aligned}\hat{a}_{1}(\theta)= & \hat{K}\left(\cosh\Omega(\hat{a}_{1}+\hat{a}_{2})+\sinh\Omega(\hat{a}_{1}^{\dagger}+\hat{a}_{2}^{\dagger})\right)\\
 & +\frac{1}{2}\hat{K}_{\Delta}\left(\cosh\Omega_{\Delta}(\hat{a}_{1}-\hat{a}_{2})+\sinh\Omega_{\Delta}(\hat{a}_{1}^{\dagger}-\hat{a}_{2}^{\dagger})\right)
\end{aligned}
\end{equation}

\begin{equation}
\begin{aligned}\hat{a}_{2}(\theta)= & \hat{K}\left(\cosh\Omega(\hat{a}_{1}+\hat{a}_{2})+\sinh\Omega(\hat{a}_{1}^{\dagger}+\hat{a}_{2}^{\dagger})\right)\\
 & -\frac{1}{2}\hat{K}_{\Delta}\left(\cosh\Omega_{\Delta}(\hat{a}_{1}-\hat{a}_{2})+\sinh\Omega_{\Delta}(\hat{a}_{1}^{\dagger}-\hat{a}_{2}^{\dagger})\right)
\end{aligned}
\end{equation}

Then, vaccume is

\begin{equation}
|\theta\rangle_{n}=e^{-\hat{K}\tanh\Omega(\hat{a}_{1}^{\dagger}+\hat{a}_{2}^{\dagger})^{2}+\frac{1}{2}\hat{K}_{\Delta}\tanh\Omega_{\Delta}(\hat{a}_{1}^{\dagger}-\hat{a}_{2}^{\dagger})^{2}}|0\rangle.
\end{equation}

Now, if we introduce

\begin{equation}
\hat{a}_{+}\equiv\hat{a}_{1}+\hat{a}_{2},
\end{equation}

\begin{equation}
\hat{a}_{-}\equiv\hat{a}_{1}-\hat{a}_{2},
\end{equation}

the vaccume become
\begin{equation}
|\theta\rangle_{n}=e^{-\hat{K}\tanh\Omega\hat{a}_{+}^{\dagger2}}e^{-\hat{K}_{\Delta}\tanh\Omega_{\Delta}\hat{a}_{-}^{\dagger2}}|0\rangle=(-K\tanh\Omega)^{n_{+}}(-K_{\Delta}\tanh\Omega_{\Delta})^{n_{-}}|n_{+}\rangle|n_{-}\rangle
\end{equation}

Now, we consider pure status matrix for $|\theta\rangle$,

\begin{equation}
\rho(\theta)=|\theta\rangle_{n}\langle\theta|.
\end{equation}

Corresponding to the fact that the canonical distribution is obtained
when one mode is integral out in the Bogoliubov-transformed vacuum,
we will now check whether that generalization is obtained in the Virasoro-Bogoliubov-transformed
vacuum. Corresponding to the decomposition of the density matrix into
center of mass and relative coordinates in deriving the Boltzmann
equation from a finite temperature field theory and integrating out
the relative coordinates as more microscopic coordinates, we integrate
out the terms that depend on the relative coordinates by integrating
out the \textquotedbl -\textquotedbl{} modes,
\begin{equation}
\rho_{+}(\theta)=\mathrm{Tr}_{-}|\theta\rangle_{n}\langle\theta|.
\end{equation}

We obtain

\begin{equation}
\rho_{+}(\theta)\propto e^{-K^{2}\tanh^{2}\Omega\hat{N}_{+}}.
\end{equation}

Then $\rho_{+}(\theta)$ is canonical distribution and $\beta$ is 

\begin{equation}
K(\theta,\langle\hat{x}^{n}\rangle)^{2}\tanh^{2}\Omega(\theta,\langle\hat{x}^{n}\rangle)=e^{-\beta}.
\end{equation}

\section{Summary and Discussion}

To extend the Vilasoro squeezing to two modes, we considered the Bogoliubov
transformation as a squeezing of the center of mass and relative coordinates,
and generalized it to Vilasoro-Squeezing for these.

As a result, a nonlinear generalization of the Bogoliubov transformation
was obtained. When one mode of the density matrix of the Bogoliubov-transformed
vacuum is integral out, it is found that the density matrix takes
a canonical distribution in the approximation taking the expectation
value. It would be interesting to explore the implications of such
a relationship for temperature when these arguments are applied to
actual physical phenomena such as Hawking radiation from a black hole.

It is also interesting to see what the statistical distribution of
the density matrix with one of the modes integral out would look like
if we do not take expectation values when obtaining the density matrix.
It is possible that Viraso-Squeezing reveals a relationship with Tsallis
statistics\cite{key-18}.

The generalization to fermionic modes is a necessary argument in extensions
of BCS theory. It is not possible to generalize the fermionic Bogoliubov
transformation to Virasoro-squeezing simply from Grassmannianity.
In this case, a generalization of the Virasoro algebra to supersymmetry
may well define a Fermion-containing Virasoro squeezing transformation.

These studies should be reserved for future research.

\section*{Acknowledgments}

I am indebted to Akio Sugamoto and Shiro Komata for reading this paper
and giving useful comments. I appreciate Yoshiki Matsuoka for reading
my paper.

\end{document}